\pdfoutput=1
\documentclass{sig-alternate}
\usepackage[frozencache=true,cachedir=.]{minted}
\usepackage{mathptmx} 

\usepackage{fancyhdr}
\usepackage[normalem]{ulem}
\usepackage[hyphens]{url}
\usepackage[sort,nocompress]{cite}
\usepackage[final]{microtype}
\usepackage[keeplastbox]{flushend}
\usepackage{hyperref}
\hypersetup{bookmarks=true,breaklinks=true,letterpaper=true,colorlinks,linkcolor=black,citecolor=blue,urlcolor=black}

\usepackage{qcircuit}
\usepackage{physics}
\usepackage{amsmath}
\usepackage{amsfonts}
\usepackage{xcolor}

\usepackage{listings}

\makeatletter
\newcommand*{\rom}[1]{\expandafter\@slowromancap\romannumeral #1@}
\makeatother

\usepackage[ruled,vlined,linesnumbered]{algorithm2e}
\SetAlgorithmName{Program}{program}{List of Programs}
\SetArgSty{textnormal}
\makeatletter
\newcommand{\removelatexerror}{\let\@latex@error\@gobble}
\newcommand{\nosemic}{\renewcommand{\@endalgocfline}{\relax}}
\newcommand{\dosemic}{\renewcommand{\@endalgocfline}{\algocf@endline}}
\let\oldnl\nl
\newcommand{\nonl}{\renewcommand{\nl}{\let\nl\oldnl}}

\makeatother

\usepackage{enumitem}

\usepackage{paralist}
\usepackage{enumerate}

\renewenvironment{itemize}[1]{\begin{compactitem}#1}{\end{compactitem}}
\renewenvironment{enumerate}[1]{\begin{compactenum}#1}{\end{compactenum}}

\usepackage{caption}
\usepackage{subcaption}

\usepackage{authblk}

\usepackage[symbol]{footmisc}

\fancypagestyle{firstpage}{
  \fancyhf{}
  
  \fancyfoot[C]{\thepage}
}

\newcommand{\sys}{QuAPE}

\title{Exploiting Different Levels of Parallelism in the Quantum Control Microarchitecture for Superconducting Qubits \vspace{-15mm}} 

\begin{document}

\author[1, \thanks{Mengyu Zhang and Lei Xie are joint first authors. Corresponding author: mengyuzhang@tencent.com} ]{Mengyu Zhang}
\author[2, $^*$]{Lei Xie}
\author[1]{Zhenxing Zhang}
\author[1]{Qiaonian Yu}
\author[1]{Guanglei Xi}
\author[1]{Hualiang Zhang}
\author[1]{\\Fuming Liu}
\author[1]{Yarui Zheng}
\author[1]{Yicong Zheng}
\author[1]{Shengyu Zhang}
\affil[1]{Tencent Quantum Laboratory, Tencent, Shenzhen, Guangdong 518507, China}
\affil[2]{Department of Computer Science and Technology, Tsinghua University, Beijing 100084, China}


\maketitle
\thispagestyle{firstpage}
\pagestyle{plain}


\begin{abstract}
   
   As current Noisy Intermediate Scale Quantum (NISQ) devices suffer from decoherence errors, any delay in the instruction execution of quantum control microarchitecture can lead to the loss of quantum information and incorrect computation results. Hence, it is crucial for the control microarchitecture to issue quantum operations to the Quantum Processing Unit (QPU) in time. As in classical microarchitecture, parallelism in quantum programs needs to be exploited for speedup. However, three challenges emerge in the quantum scenario:  1) the quantum feedback control can introduce significant pipeline stall latency; 2) timing control is required for all quantum operations; 3) QPU requires a deterministic operation supply to prevent the accumulation of quantum errors. 
   
  In this paper, we propose a novel control microarchitecture design to exploit Circuit Level Parallelism (CLP) and Quantum Operation Level Parallelism (QOLP). Firstly, we develop a \emph{Multiprocessor} architecture to exploit CLP, which supports dynamic scheduling of different sub-circuits. This architecture can handle parallel feedback control and minimize the potential overhead that disrupts the timing control. Secondly, we propose a \emph{Quantum Superscalar} approach that exploits QOLP by efficiently executing massive quantum instructions in parallel. Both methods issue quantum operations to QPU deterministically. In the benchmark test of a Shor syndrome measurement, a six-core implementation of our proposal achieves up to 2.59$\times$ speedup compared with a single core.  For various canonical quantum computing algorithms, our superscalar approach achieves an average of 4.04$\times$ improvement over a baseline design. Finally, We perform a simultaneous randomized benchmarking (simRB) experiment on a real QPU using the proposed microarchitecture for validation.
  


\end{abstract}

\section{Introduction}

    

\begin{figure}[t]
\includegraphics[width=0.45\textwidth]{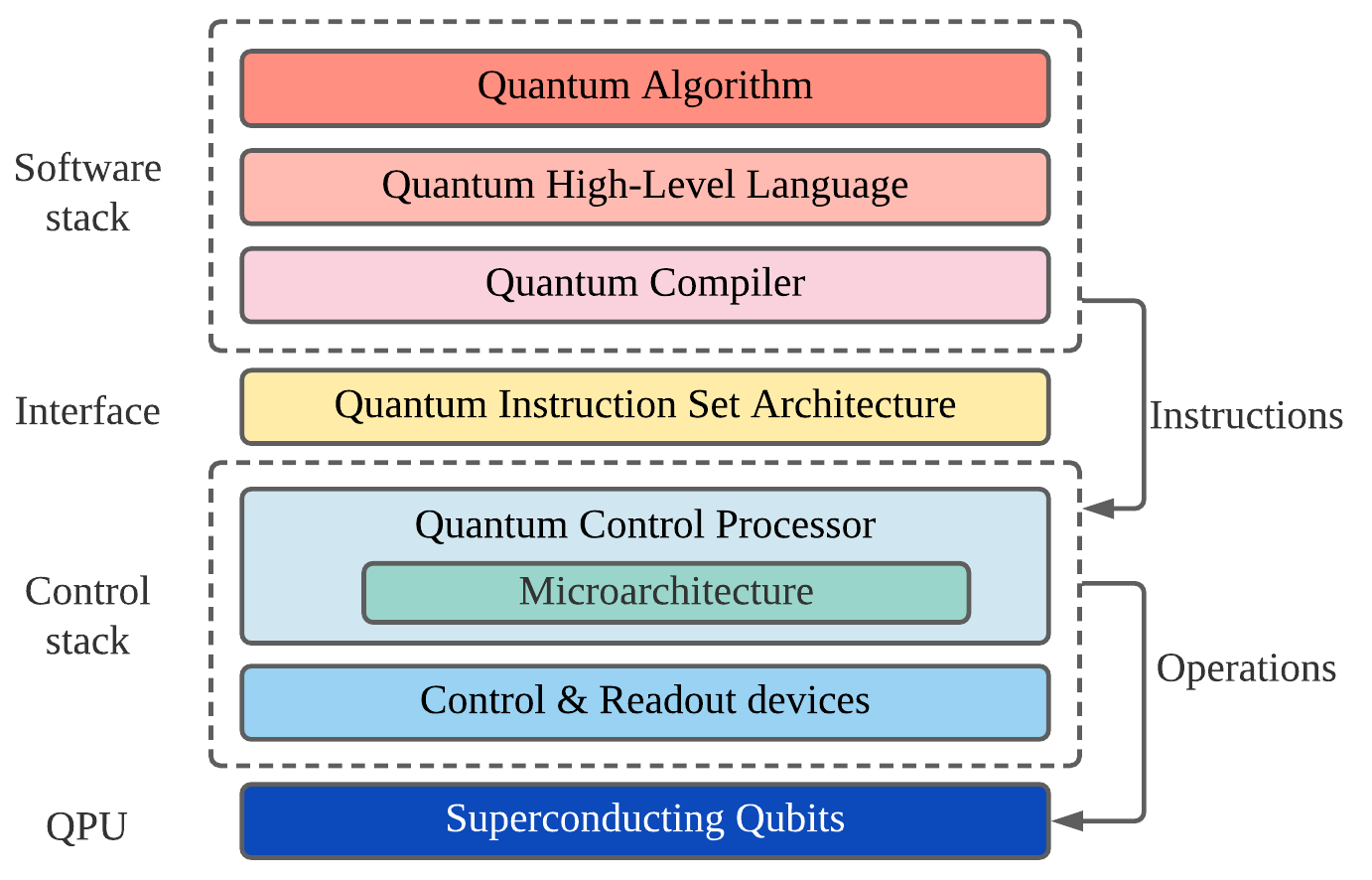}
\centering
\setlength{\abovecaptionskip}{3pt}
\caption{Organization of a full stack superconducting quantum computer. The QISA layer acts as an interface between quantum software and hardware. 
The control microarchitecture is implemented in the control processor. The control stack is composed of a control processor and analog devices, which issue quantum operations to the quantum processor.
\vspace{-5mm}}
\label{fig: full stack}
\end{figure}

    As quantum algorithms for solving problems such as integer factorization \cite{shor1994algorithms} are far beyond the reach of near-term quantum processors, building a fully-programmable quantum computer with the Noisy Intermediate Scale Quantum (NISQ) computing model \cite{preskill2018quantum} becomes the near-term goal. To this end, the quantum software and hardware need to work seamlessly. Existing high-level quantum software can provide different types of control flow to quantum algorithms \cite{cross2017open, svore2018q, chong2017programming}, but they are not readily to be implemented by hardware yet. Like in classical computer systems where between software and chip lie the instruction set architecture and microarchitecture, quantum instruction set architecture (QISA) \cite{smith2016practical, fu2019eqasm} and control microarchitecture designs \cite{fu2017experimental} have been proposed to bridge the gap between quantum software and hardware. A full-stack quantum computer is illustrated as in Figure \ref{fig: full stack}, where the post-compilation instructions are executed by the control microarchitecture to issue quantum operations to the quantum processing unit (QPU).

    


Unlike conventional systems, an inefficient quantum control microarchitecture not only affects the speed, but may also introduce decoherence errors and make the computation result incorrect. 
In current NISQ devices, any delay in quantum operations issued from the microarchitecture can result in additional accumulated quantum errors. As much damage as this can incur, we also note that this type of delay is mainly caused by the limited operation issue rate of the control microarchitecture, which can be significantly alleviated by exploiting parallelism in the control microarchitecture.


As in classical programs, quantum programs also have some natural parallelism opportunities at different levels, with two prominent ones as follows. (1) Many quantum programs are logically structured as parallel sub-circuits to be executed. (2) In each sub-circuit, different quantum operations can be executed simultaneously. 
These pre-defined parallelism needs to be properly exploited by the control microarchitecture to avoid additional delays in issuing operations to the QPU.
In this paper, we define these two levels of parallelism as follows.


\noindent (\rom{1}) \textbf{Circuit Level Parallelism (CLP)}: Quantum circuits for relevant applications may well contain sub-circuits that can be executed in parallel to certain extent. An ideal control microarchitecture should be capable to exploit this parallelism and support \textit{parallel processing} of the sub-circuits.

    
\noindent (\rom{2}) \textbf{Quantum Operation Level Parallelism (QOLP)}: Operations issued by control microarchitecture can be executed simultaneously without affecting the semantics, and an ideal control microarchitecture should be capable to exploit this parallelism and support issuing quantum operations to different qubits in parallel within a certain period of time. To estimate the performance of QOLP exploitation, we propose \textit{Cycles Each Step} (CES) and \textit{Time Ratio} (TR) metrics that measure the execution time of classical control part and quantum execution part.


    
    \vspace{3mm}
    
Previous research has shown that some quantum experiments with only several qubits already require nontrivial effort to deliver quantum operations \cite{fu2019eqasm} due to limited speed of the control microarchitecture. In the current NISQ era with dozens or even hundreds of noisy qubits, it is more important to exploit CLP and QOLP to reduce the delay in issuing operations.



To our knowledge, no solution has been proposed to directly exploit CLP. Regarding QOLP, Fu \textit{et al.} mainly provided solutions to improve the instruction information density at the QISA level \cite{fu2019eqasm}, but there still lacks a microarchitecture-level method that can efficiently and effectively exploit QOLP during run-time. 

    
    
Many classical solutions for parallelism exploitation exists, such as multiple-issue and speculation \cite{hennessy2011computer}. However, these solutions cannot be directly applied to quantum microarchitecture. Due to distinctive quantum features, the quantum microarchitecture design needs to solve the following problems.

    
    \noindent (1) \textbf{Quantum feedback control} is a special control flow for the quantum scenario.
    This type of control requires real-time interaction between QCP and QPU.
    It refers to intermediate measurements, and branching according to the measurement outcomes, in a quantum circuit. This feedback control significantly complicates the parallelism exploitation by introducing pipeline stalls. Moreover, it hinders the CLP exploitation when occurred simultaneously.
    
    
    \noindent (2) \textbf{Timing control} is required in the control microarchitecture to issue quantum operations with accurate timing \cite{fu2017experimental}. This mechanism brings timing dependency for all quantum instructions, which is a different type of dependency than auxiliary classical instructions. Hence, the timing control forms a hurdle for parallelism exploitation. 
    
    

    \noindent (3) \textbf{Deterministic operation delivery.}
    The QPU requires a deterministic supply of quantum operations \cite{tannu2017taming}.
    Indeed, many classical techniques such as branch prediction cannot be directly translated into quantum microarchitecture, because the prediction failure can lead to substantial amount of errors given the short coherence time of superconducting qubits \cite{koch2007charge}.
    
    
    To address these challenges, we propose a practical microarchitecture design, named 
    \sys{} (Quantum control microArchitecture for Parallelism Exploitation)
    for superconducting qubits to exploit different levels of parallelism. Overall, our contributions in this work are:
    
    \begin{enumerate}
        \item To exploit CLP, we design a \textit{Multiprocessor} architecture that supports parallel processing of feedback control by allocating different sub-circuits to multiple processing units. The mechanisms can significantly reduce potential overhead that hinders the timing control of the quantum program.
        \item To exploit QOLP, we propose a \textit{Quantum Superscalar} method which is capable of issuing a large amount of quantum operations to the QPU in a deterministic manner. We also introduce a method to absorb the stall latency caused by simple feedback controls.
    \end{enumerate}
    
    A prototype of \sys{} is implemented on a field programmable gate array (FPGA), and various experiments are conducted for performance evaluation. In a benchmark test of a Shor syndrome measurement, our multiprocessor architecture achieves up to 2.59$\times$ speedup in the execution time. We also evaluate the performance of our superscalar approach on several benchmarks from ScaffCC \cite{scaffcc}, Qiskit \cite{Qiskit} and RevLib \cite{WGT+:2008}, and observe an average improvement of 4.04$\times$ in TR. We validate our design by performing a simultaneous randomized benchmarking (simRB) \cite{gambetta2012characterization} experiment on a superconducting QPU, showing that our design is capable of simultaneously apply quantum gates to different qubits. 
    
\section{Background}


    

\subsection{Quantum Instruction Set Architecture}
    
    Similar to classical architectures \cite{hennessy2011computer}, the QISA is an essential component of a fully programmable quantum computer and acts as an interface between the quantum software and hardware.
    In this work, we target an executable QISA for current NISQ hardware which supports the following two features \cite{fu2019eqasm}: 
    
    \begin{itemize}
        \item It provides explicit timing information to help the control microarchitecture to achieve timing control of the issued quantum operations (see Section 2.2). 
        \item In addition to quantum instructions, it provides auxiliary classical instructions to provide control flow, such as loops and feedback control.
    \end{itemize}

\subsection{Quantum Control Processor}
    
    
    The control microarchitecture implemented in the Quantum Control Processor (QCP) accepts post-compilation instructions as input. 
    These instructions can be classified into two main types: (1) \textbf{classical instructions}, mainly used for constructing different types of control flow, and usually consisting of four kinds of instructions: control, data transfer, logical, and arithmetic; (2) \textbf{quantum instructions} that describe quantum operations. These quantum instructions are executed in the QCP to issue corresponding quantum operations to the QPU. 
    Here, we emphasize the difference between \textbf{quantum instructions} and \textbf{quantum operations}: quantum instructions are executed on QCP, while quantum operations are executed on QPU. To avoid ambiguity, ``\textbf{issue}'' is only used to indicate the procedure of QCP sending quantum operations into QPU in this work.
    
    To achieve timing control of the issued quantum operations, timing labels are added to the quantum instructions, which represents the time interval since the issue of the quantum operation corresponding to the previous quantum instruction. For example, the following assembly code represents an \texttt{H} operation applied on \texttt{q0} and also on \texttt{q1} at the same time, and a \texttt{CNOT} operation for these two qubits in the next step. 
    The number at the beginning of each line can be regarded as the timing label of each instruction. The timeline is constructed by executing quantum instructions in QCP. The QCP can issue quantum operations with accurate timing on nanosecond scales. 
    We refer the interested reader to the original papers for a detailed introduction \cite{fu2017experimental, fu2019eqasm}.

{\renewcommand\fcolorbox[4][]{\textcolor{cyan}{\strut#4}}
\begin{minted}[linenos,frame=single]{nasm}
0 H      q0
0 H      q1
1 CNOT   q0, q1
\end{minted}
}
\vspace{-3mm}

\subsection{Superconducting Qubits}

    There are various physical platforms for implementing qubits \cite{kelly2015state, debnath2016demonstration, hanson2007spins, cramer2016repeated}. Among them, the Superconducting Qubits (SQ) are regarded as one of the front-runner technologies due to its comparatively high scalability and gate fidelity. This paper focuses on QPUs for superconducting qubits.
    
    The relatively short coherence time of SQ (50-100 $\mu$s) gives a huge challenge for classical control. For example, the feedback control for quantum error correction \cite{lidar2013quantum} needs to be completed within 1\% of this coherence time to achieve the fault-tolerance  \cite{fowler2012towards,tomita2014low}. The typical gate time for single-qubit operation and two-qubit operation are 20ns and 40ns, respectively. The measurement operation usually requires a pulse of 100ns - 2$\mu$s.
    

\subsection{Feedback Control}

    
    Unlike static quantum circuits that use a predetermined list of sequences, dynamic circuits rely on conditional logic to respond to measurement results in real time. In the quantum scenario, the \emph{feedback control} is a distinctive control flow, which can be divided into the following stages:
    
    \begin{enumerate}[(I)]
      \item The QCP executes the measurement instruction and starts to wait for the result to return. The measurement operation is performed on the QPU.
      \item The Digital Acquisition (DAQ) receives analog results and performs digital acquisition.
      \item The QCP obtains the classical state and begins to execute corresponding conditional logic.
      \item The determined quantum operation is issued to the QPU.
    \end{enumerate}

\begin{figure}[h]
\centering
\includegraphics[width=0.48\textwidth]{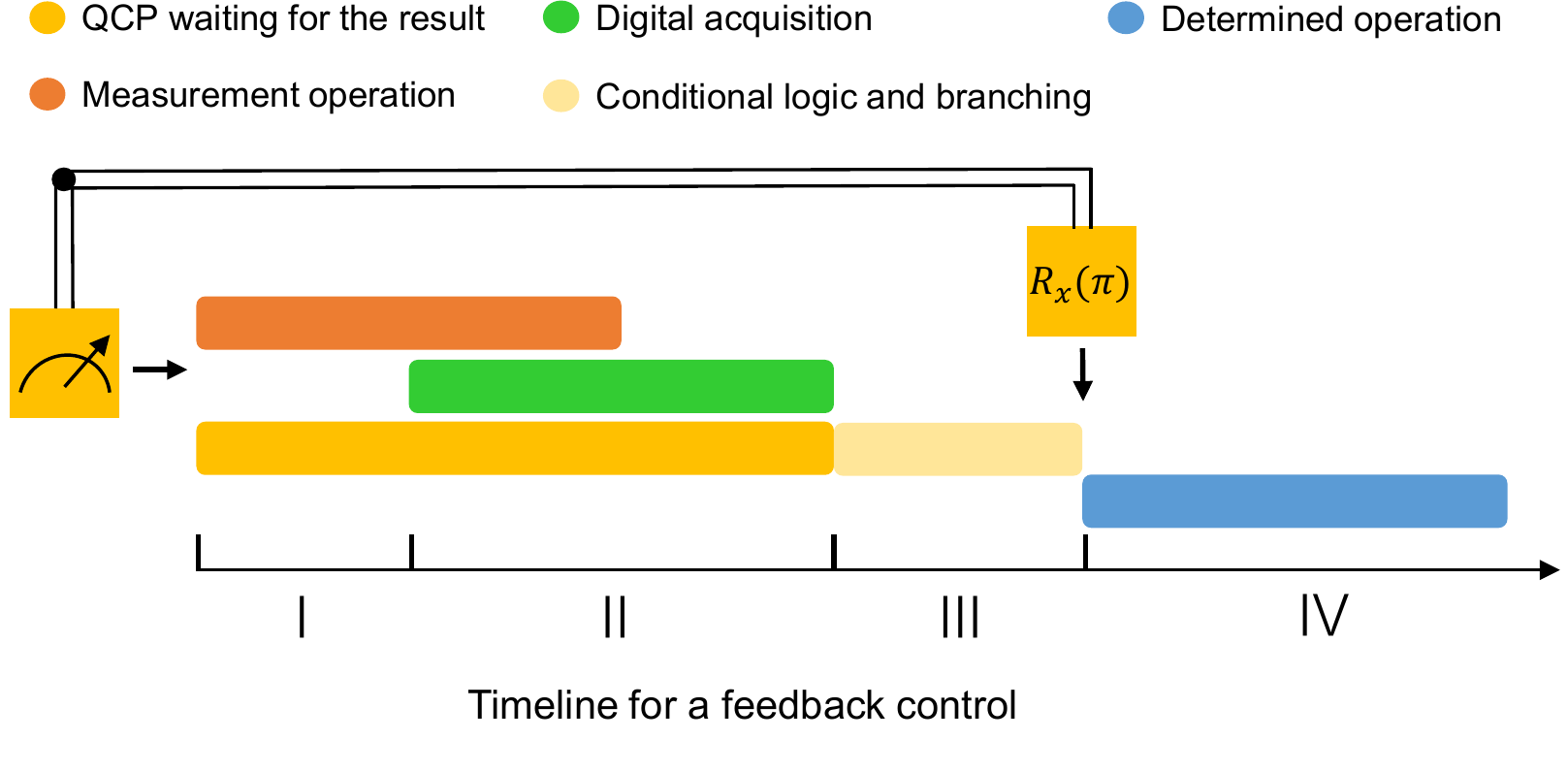}
\caption{The latency breakdown for a feedback control process to determine whether a $R_x(\pi)$ gate needs to be applied. The arrows indicate the time when the QCP executes the quantum instruction. \vspace{-3mm}}
\label{fig:fed ctrl}
\end{figure}
    

    Figure \ref{fig:fed ctrl} outlines the timeline of a quantum dynamic circuit. The latency of Stage \rom{1} and \rom{2} are non-deterministic due to the intrinsic uncertainty of dispersive readout \cite{wallraff2004strong}. To avoid reading an invalid result, previous research implemented a synchronization protocol to stall the pipeline until the measurement result is ready \cite{fu2019eqasm}. The time spent in Stage \rom{3} comes from the QCP execution delay of conditional logic and branching, which depends on the complexity of the classical operations. The overall latency of feedback control typically requires at least hundreds of nanoseconds.
    
    With feedback control, the QCP can support a wide range of dynamic quantum circuits, including active qubit reset \cite{riste2012feedback}, quantum teleportation \cite{bennett1993teleporting,bouwmeester1997experimental}, and iterative phase estimation \cite{kitaev1995quantum,kitaev2002classical}. Repeat-Until-Success (RUS) \cite{paetznick2013repeat} is a special case of dynamic quantum circuits, which repeatedly execute certain quantum operations and perform a measurement until a success measurement outcome occurs. Such circuits bring uncertainty to the program execution time, and therefore needs a careful treatment.

\section{Two Levels of Parallelism}
    
    
    In this section, we give an overview of the two levels of parallelism of quantum circuits. 

\subsection{Circuit Level Parallelism}
    
    CLP is defined as the form of parallelism among different sub-circuits or circuits. We use the term \textbf{program block} in this paper to refer to a sequence of instructions containing quantum instructions and control flow of the corresponding sub-circuit. The CLP can be found in the following two different scenarios. We use an example to study the impact of CLP exploitation on microarchitecture design. 
    

\subsubsection{Scenario 1: Parallel processing}
    
    A typical case for parallel processing is a single quantum application that needs to be partitioned into multiple \emph{parallel} sub-circuits. For complex quantum applications, these divided program blocks may contain feedback control and loops, whose CLP needs to be exploited. An example of such application is Shor syndrome measurement (see Section 7).
        
        
        
\subsubsection{Scenario 2: Multiprogramming}    

    This situation describes multiple tasks that are relatively independent and supposed to be executed on the same QPU simultaneously. For example, the QCP needs to exploit CLP to run multiple programs generated through multiprogramming, especially when these tasks represent dynamic circuits. This method helps to improve the resource utilization of quantum cloud services and is discussed in \cite{das2019case} at the software level.

\subsubsection{Example}
    
\begin{figure}[h]
\centering
\includegraphics[width=0.49\textwidth]{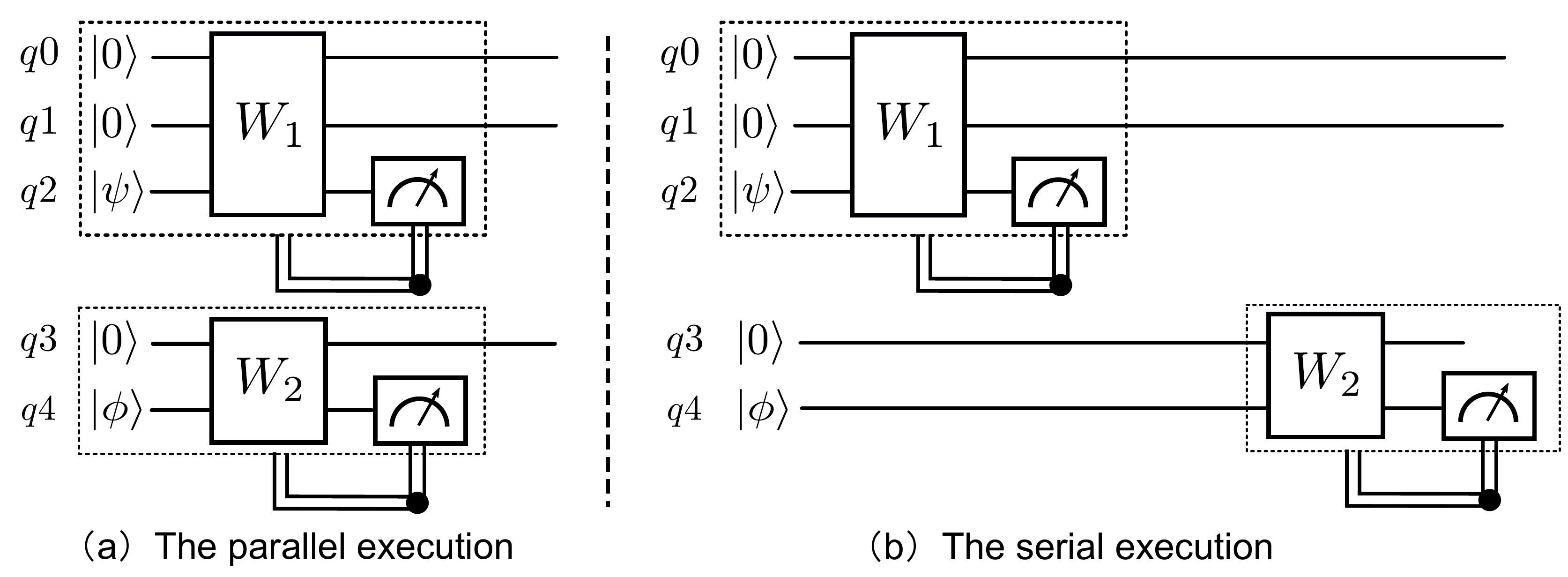}
\setlength{\abovecaptionskip}{2pt}
\caption{Different execution results of two parallel repeat-until-success sub-circuits.}
\label{fig:RUS}
\end{figure}

    We use a circuit with two parallel RUS sub-circuits as an example to illustrate the challenge for exploiting CLP. Figure \ref{fig:RUS}(a) shows ideal execution of this circuit on the QPU. The assembly program adapted to the previous design requires using one control flow to describe the circuit, as written by Program \ref{algo:reorgranize}.
    
    \begin{algorithm}[h]
        \DontPrintSemicolon
        \SetKwProg{subproc}{Procedure}{}{}
        \caption{Parallel RUS Circuits within one control flow}
        \label{algo:reorgranize}

        $Branch 1:$ operations of $W_1$, $W_2$, measure ancilla qubits $\texttt{q}2$ and $\texttt{q}4$ \;
        $Branch 2:$ operations of $W_1$, measure ancilla qubits $\texttt{q}2$\;
        $Branch 3:$ operations of $W_2$, measure ancilla qubit $\texttt{q}4$\;
        $Branch 4:$ Continue\;
        \uIf{Both fails}{ 
            perform correction and reset on both sub-circuits \; 
            Go to $Branch 1$
        }
        \uElseIf{$W_1$ fails, $W_2$ success}{
            perform correction and reset on $W_1$ \; 
            Go to $Branch 2$
        } 
        \uElseIf{$W_2$ fails, $W_1$ success}{
            perform correction and reset on $W_2$ \; 
            Go to $Branch 3$
        }
        \Else{
            Go to $Branch 4$
        }
    \end{algorithm}
    
    
    
    Using one feedback control flow to describe all sub-circuits introduces $O(2^n)$ ($n$ is the total number of sub-circuits) branches. Each branch contains the quantum operations of the entire circuit, which causes additional execution delay and program size. Obviously, this method is sophisticated and not scalable. Therefore, a feasible method is to use independent program blocks to describe different sub-circuits, which only require $O(2n)$ branches. The corresponding program is given in Program \ref{algo:circuit_schedule}.
    \begin{algorithm}[h]
        \DontPrintSemicolon
        \SetKwProg{subproc}{Procedure}{}{}
        \caption{Parallel RUS Circuits in two blocks}
        \label{algo:circuit_schedule}

        $W_1:$ perform quantum operations of sub-circuit $W_1$\;
        $Meas:$ measure ancilla qubits $\texttt{q}2$\; \label{line:exe}
        \eIf{measurement results represent a failure}{ \label{line:type_start}
            perform correction operation on $\texttt{q}2$ \; \label{line:single}
            perform reset on $\texttt{q}0$ and $\texttt{q}1$ \;
            Go back to $W_1$
        }{
            Continue \;
        } \label{line:type_end}

        \nonl \;

        $W_2:$ perform operations of $W_2$\;
        $Meas:$ measure $\texttt{q}4$\; \label{line:exe}
        \eIf{failure}{ \label{line:type_start}
            perform correction and reset\;
            Go back to $W_2$
        }{
            Continue \;
        } \label{line:type_end}
    \end{algorithm}
    
    
    However, the control microarchitecture with uniprocessor cannot handle such parallel program blocks. The QCP will not execute any instruction from sub-circuit $W_2$ before the termination of the first program block, because an executing program ``blocks'' the processing of other ones. In this case, the ideal parallel execution is forced into a serial execution (Figure \ref{fig:RUS}(b)). 
    This situation is unacceptable because the latency of the entire sub-circuit $W_1$ is added to all other qubits. Hence, it is an open challenge to design microarchitecture-level methods for CLP exploitation under the structure of Program \ref{algo:circuit_schedule}.

\subsection{Quantum Operation Level Parallelism}
    
    Quantum Operation Level Parallelism (QOLP) is used to measure a different level of parallelism other than CLP. The parallel quantum operations can be defined as operations that start simultaneously on different qubits, regardless of the duration time of operations. To quantify the QOLP exploitation, we define \textit{cycles each step} (CES) and TR in the following.
    

\subsubsection{Cycles each step}

\begin{figure}[h]
\centering
\includegraphics[width=0.3\textwidth]{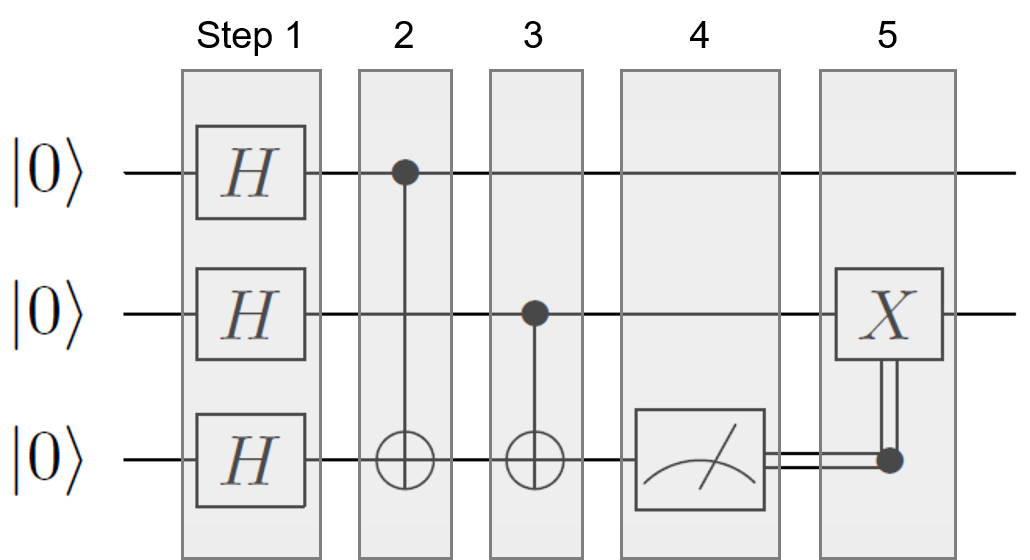}
\setlength{\abovecaptionskip}{3pt}
\caption{The circuit for deterministic entanglement creation with a feed-forward logic. The grey box indicates different steps in this circuit. \vspace{-5mm}}
\label{fig:circuit_step}
\end{figure}

    We define circuit step as a component of quantum circuit, which contains all parallel quantum operations at a certain timing point. For example, we can divide the circuit for deterministic entanglement creation into five steps, as shown in Figure \ref{fig:circuit_step}. Different circuit steps can ensure the execution order of the quantum program. 

    To measure the QCP performance on QOLP, we define \textit{cycles each step} (CES) as the number of clock cycles required by QCP to execute instructions for each circuit step. The value of the CES is the sum of the following four parts: quantum instruction execution cycles, classical instruction cycles, control stalls caused by loops and subroutines, and the contribution of feedback control. The ideal \textit{cycles each quantum instruction} (CEQI) is 1 in a pipelined processor, and \textit{quantum instruction count each step} (QICES) is the number of quantum instructions contained in one circuit step of the program. The CES is defined in Equation (\ref{eq:CES}).
    \begin{equation}
    \begin{aligned}
        CES = & \hspace{0.1em} (pipeline \hspace{0.4em} CEQI) \times QICES \\ &+ classical \hspace{0.4em} instruction \hspace{0.4em} cycles\\&+ classical \hspace{0.4em} control \hspace{0.4em} stalls \\ &+ QCP \hspace{0.4em} execution \hspace{0.4em} delay \hspace{0.4em} of \hspace{0.4em} feedback \hspace{0.4em} control
    \end{aligned}
    \label{eq:CES}
    \end{equation}
    
    For example, step 1-4 in Figure \ref{fig:circuit_step} only contain the quantum instruction execution part. Since step 5 involves a feedback control, it must wait for a period of time as explained in Section 2.5. The delay of Stage \rom{1} and \rom{2} is unavoidable for both QCP and QPU, and is not calculated in the CES. The fourth part of the CES comes from the conditional execution in the QCP, which is equivalent to the delay of Stage \rom{3}.
    

\subsubsection{Time ratio}
    
    The QCP should ensure that the processing of all quantum instructions in a single circuit step is completed within the execution time spent on the QPU. Quantifying this target, we define \textit{Time Ratio} (TR) as the execution time of QCP compared to the execution time of QPU. The TR of step $i$ is defined in Equation (\ref{eq:ratio}). 
    In each circuit step, the QPU executes quantum operations in fully parallel, allowing us to use the quantum gate time to represent the QPU time.
    \begin{equation}
    TR_i = \frac{QCP \hspace{0.4em} time \hspace{0.4em} step \hspace{0.4em} i}{QPU \hspace{0.4em} time \hspace{0.4em} step \hspace{0.4em} i}=\frac{clock \hspace{0.4em}  time \times CES_i}{gate \hspace{0.4em} time}
    \label{eq:ratio}
    \end{equation}
    
    Since even a small delay of quantum operations can result in the loss of quantum states and incorrect execution result, the goal of QOLP exploitation is to reach $\mathbf{TR \leq 1}$ for the entire program. This requirement poses a challenge for the control microarchitecture design, because CES will grow rapidly with the number of qubits while the gate time remain unchanged.
    
    Since the parallelism of sub-circuits will cause another level of latency, we only use CES and TR to evaluate the performance of QOLP exploitation.
    

\section{Requirements}
    
    This section introduces specific requirements for microarchitecture stemmed from quantum computing. We also discuss the design guidelines for tackling these challenges during parallelism exploitation.

\subsection{Feedback control}

    
    Feedback control occurs frequently in a range of quantum applications. Feedback control in quantum programs contribute to CES as shown in the later two components in Equation \ref{eq:CES}. Even worse, these control flow hinder the CLP exploitation when occurred simultaneously (see Section 3.1.3). Moreover, the non-determinism in some special feedback control (e.g., RUS) can lead to additional execution overhead during parallelism exploitation. To tackle the problems caused by feedback control, the microarchitecture needs to (1) absorb the feedback control latency, (2) manage parallel feedback control, and (3) eliminate potential overhead caused by non-deterministic circuits. 

\subsection{Timing Control}

    
    The timing control ensures accurate timing of issued quantum operations by adding pre-determined timing labels to quantum instructions (see Section 2.3). To comply with the timing control, the QCP needs to preserve timing dependency of different quantum instructions during parallelism exploitation. Moreover, the timing of operations can significantly impact the fidelity of the final results when the pre-scheduled timing control is disrupted by unexpected overhead. Therefore, the QCP should eliminate potential overhead caused by the parallelism exploitation methods.
    
    

\subsection{Deterministic Operation Supply}


    Solutions such as branch prediction are widely used to improve the performance of classical computer architectures. However, this kind of method is non-deterministic, which will bring a corresponding penalty when the failure occurs. Such non-determinism makes it more difficult to meet the requirement of $TR \leq 1$ for the entire program. To minimize the uncertainty of classical control, the QCP needs to use deterministic techniques to exploit CLP and QOLP. 
    
    
    
    

\begin{figure*}[t]
\centering
\includegraphics[width=0.85\textwidth]{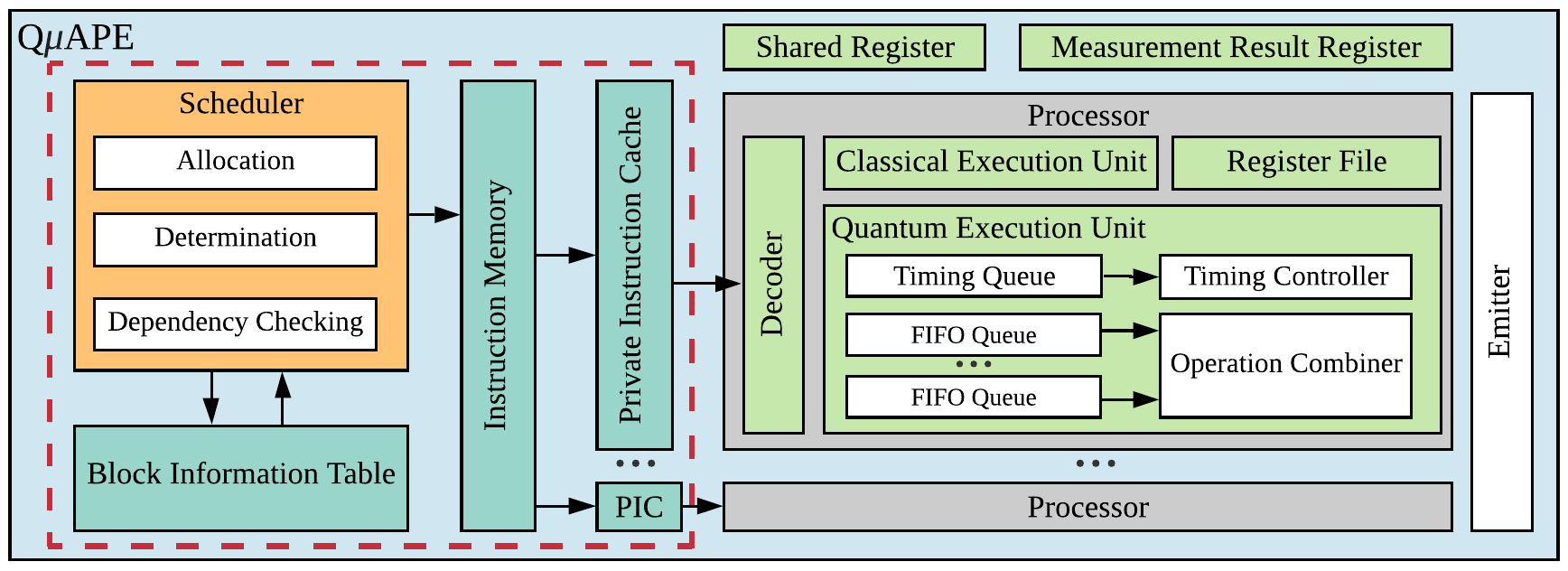}
\setlength{\abovecaptionskip}{3pt}
\caption{Quantum control microarchitecture implementing the multiprocessor architecture. The grey dotted box represents the control unit for determining the collaboration among multiple processors. \vspace{-4mm}}
\label{fig:multiprocessor}
\end{figure*}

\section{Microarchitecture}

    
    In this section, we first give an overview of the \sys{} design and then dig into the details of the structure and each component. We choose a centralized-memory architecture as the starting point. A timed QASM is chosen as the instruction set due to its explicit timing definition. \sys{} outputs signals used to control analog devices, which finally issues quantum operations for the QPU.
    

\subsection{Design Motivation}

    
    \noindent \textbf{CLP exploitation:} the CLP cannot be exploited by merely increasing the instruction execution speed of the microarchitecture, because various control will hinder the parallel execution of instructions. For instance, the program describing circuits like RUS needs to be structured as a combination of subroutine and feedback control (see Program 2), because the entire circuit is non-deterministic. The stall caused by feedback control is catastrophic for irrelevant qubits due to the long delay. A feasible method to manage parallel control flow is to process multiple program blocks concurrently. Therefore, we adopt a multiprocessor architecture to exploit CLP in the control microarchitecture.
    This architecture support dynamic scheduling of different program blocks during run-time. We also introduce a mechanism to reduce the potential overhead caused by block switching. 
    
    \noindent \textbf{QOLP exploitation:} the contribution of classical instructions to CES mainly comes from control stalls, rather than the instruction execution. The parallel execution of classical instructions is less attractive because it introduces additional complexity but does not effectively reduce CES. Conversely, the parallel execution of quantum instructions in the QCP will not encounter data hazards that are difficult to handle in classical processors. Therefore, the use of multiple-issue mechanisms can effectively eliminate the increase in QICES. To meet the requirement of TR, we adopt a quantum superscalar architecture to exploit QOLP in the microarchitecture.
    We implement a specific instruction scheduling method to prevent introducing excessive hardware complexity. This approach allows separate dispatch of quantum instructions and classical instructions, which can help to absorb the control stall latency caused by conditional branching.
    
    \noindent \textbf{Simple feedback control:} the contribution of feedback control in the CES can also be reduced. In addition to complex circuit-level feedback control such as RUS, there are also many application that only require simple feedback control e.g., active qubit reset \cite{riste2012feedback}. As describing this type of control does not require a large number of instructions, we implement a fast context switch mechanism to eliminate the feedback control stall latency.
    
    
    
    
    \vspace{1.0mm}
    
    Overall, three solutions are introduced in \sys{}. \textit{a)} The multiprocessor architecture features multiple processors with identical access to a centralized shared-memory. \textit{b)} The quantum superscalar fetches multiple instructions in one cycle and performs pre-decoding to determine instruction dispatching. 
    \textit{c)} The context switch scheme for simple feedback control uses a series of register to store control information, thereby allowing \sys{} to continue the execution before the result is returned. Scheme \textit{a)} is used to exploit CLP, while the QOLP exploitation is addressed by \textit{b)} and \textit{c)}.

\subsection{Multiprocessor}
    
    The basic structure of the proposed multiprocessor is shown in Figure \ref{fig:multiprocessor}. In this approach, all instructions generated by a quantum program are stored in the instruction memory and shared by all processing units. The post-compilation information about the partition of program blocks are stored in a block information table. Each processing unit has its own instruction cache, program counter, and execution unit. 
    
    The scheduling of \textit{multiprogramming} can be easily achieved by pre-determining the allocation of different tasks. In contrast, the \textit{parallel processing} of partitioned sub-circuits needs to absorb potential overhead caused by the uncertainty of program execution. To this end, a dynamic scheduling method is introduced. To reduce the overhead caused by program block switching, a mechanism for prefetching the instructions into private instruction caches is implemented.

\subsubsection{Block information table}
    
\begin{figure}[h]
\centering
\includegraphics[width=0.3\textwidth]{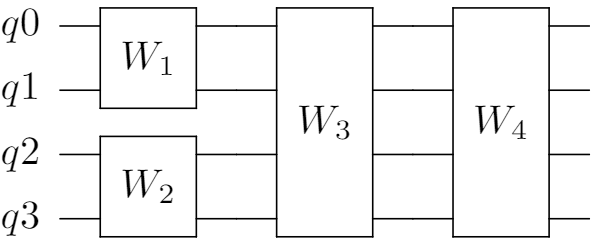}
\setlength{\abovecaptionskip}{5pt}
\caption{An example of a quantum circuit that consists of four sub-circuits.}
\label{fig:W4}
\end{figure}

    Information about each program block in the quantum program is stored in the block information table before the processing starts. Since the execution time of quantum circuits like RUS cannot be predetermined, we use the dependency of different program blocks to ensure the execution order of the quantum program. For example, the quantum circuit of Figure \ref{fig:W4} is divided into four sub-circuits and the information of corresponding program block is stored in Table \ref{table:cit}. The information includes the number of each program block, the address of the corresponding instructions in the main memory, and the dependency relationship. The dependency information of $W_1$ and $W_2$ shows that they can be executed in parallel immediately, while $W_3$ and $W_4$ need to wait for the end of the previous stage. The information in this table is read by the scheduler when a program block can be allocated to a processor.

\begin{scriptsize}
\begin{table}[h!]
  \centering
  \begin{tabular}{|l|l|l|l|}
    \hline
    \textbf{Program} & \textbf{PC start} & \textbf{PC end} & \\
    \textbf{block} & \textbf{address} & \textbf{address} & \textbf{Dependency}\\
    \hline
    \hline
    $W_1$ & 0 & 10 & None \\
    \hline
    $W_2$ & 11 & 20 & None\\
    \hline
    $W_3$ & 21 & 30 & $W_1, W_2$\\
    \hline
    $W_4$ & 31 & 40 & $W_3$\\
    \hline
  \end{tabular}
  \setlength{\abovecaptionskip}{5pt}
  \caption{An example of the block information table content of the circuit shown in Figure \ref{fig:W4}. \vspace{-4mm}}
  \label{table:cit}
\end{table}
\end{scriptsize}
    
    

\subsubsection{Scheduler}
    
    The scheduler will continuously read the values in the block information table when microarchitecture starts to run. The first stage of scheduling is to perform a dependency check on the returned block information. By dynamically checking the status of each processor, the scheduler can determine the allocation of program blocks during run-time.

    The most straightforward way to represent the dependency is to use direct addressing of all program blocks. This representation requires using bit vectors with the same bit width as the number of circuit blocks.
    For example, the dependency of block $W_3$ is the address of the first two blocks, while $W_4$ records the address of $W_3$. In this scheme, the program block is ready to be executed when its dependency is zero. When the execution of $W_1$ and $W_2$ are finished, the lower two bits in the dependency information of the entire table are cleared. Thus, $W_3$ can pass the dependency check and start the next step. Since the parallelism of program blocks is not pre-determined, direct correlation provides substantial space for dynamic scheduling.
    
    
    
    However, the direct dependency becomes memory-consuming when the number of blocks scale up. Therefore, we provide a relatively simple method to indicate the dependency by assigning a priority to each block. A program block with a higher priority means that it needs to be executed first, while blocks with the same priority signify potential parallelism. The scheduler uses a priority counter for dependency check. The counter is incremented once the execution of all blocks with the old priority is finished. The program blocks that its priority equal to the counter can pass the dependency check. 
    
    \begin{scriptsize}
    \begin{table}[h!]
      \centering
      \begin{tabular}{|l|l|l|l|l|}
        \hline
        Program block & $W_1$ & $W_2$ & $W_3$ & $W_4$ \\
        \hline
        Priority & 0 & 0 & 1 & 2 \\
        \hline
      \end{tabular}
      \label{table:dependency2}
    \end{table}
    \end{scriptsize}
    
    
    
    The compiler can choose the appropriate representation for different scenarios to exploit the CLP in quantum programs. Regardless of which representation method is chosen, the modules after the dependency check are the same. The results of the dependency check are then used to determine the allocation of each program block to the processor. The non-deterministic quantum circuits like RUS lead to an uncertain program execution time. In this case, unexpected overhead can be caused when the allocation is not effectively scheduled. To prevent such situation, we use a dynamic scheduling method to determine the program block allocation during run-time. 
    
    A series of status registers are used to record the status of each program block, including \textit{wait}, \textit{in execution}, and \textit{done}. The initial state of all blocks is "wait". When the block passes the dependency check, it starts to request for allocation. its status can be changed to "in execution" when there is an idle processor, and the corresponding instructions are fetched into the private instruction cache. During allocation, the scheduler is busy and do not answer to other requests. The processor will return a signal to the scheduler when the execution ends. The state of the corresponding block is then changed to "done", which also indicates the scheduling of subsequent blocks. With this method, the program block allocation can be finished in an efficient way.

\subsubsection{Private instruction cache}

    
    \noindent \textbf{Prefetching:} During allocation, the instructions in the program block will be fetched from the centralized main memory to the private instruction cache of each processor. The processor needs to switch to the next program after the current execution is complete. It takes certain time for the scheduler to fetch new instructions into the private cache, which may exceed the expected time for the next quantum operation to start acting on the QPU. In order to prevent the computational process from additional accumulated quantum errors, we implement a \textbf{prefetching} mechanism to minimize the overhead caused by block switching.
    
\begin{figure}[h]
    \begin{subfigure}[b]{0.48\textwidth}
         \centering
         \includegraphics[width=\textwidth]{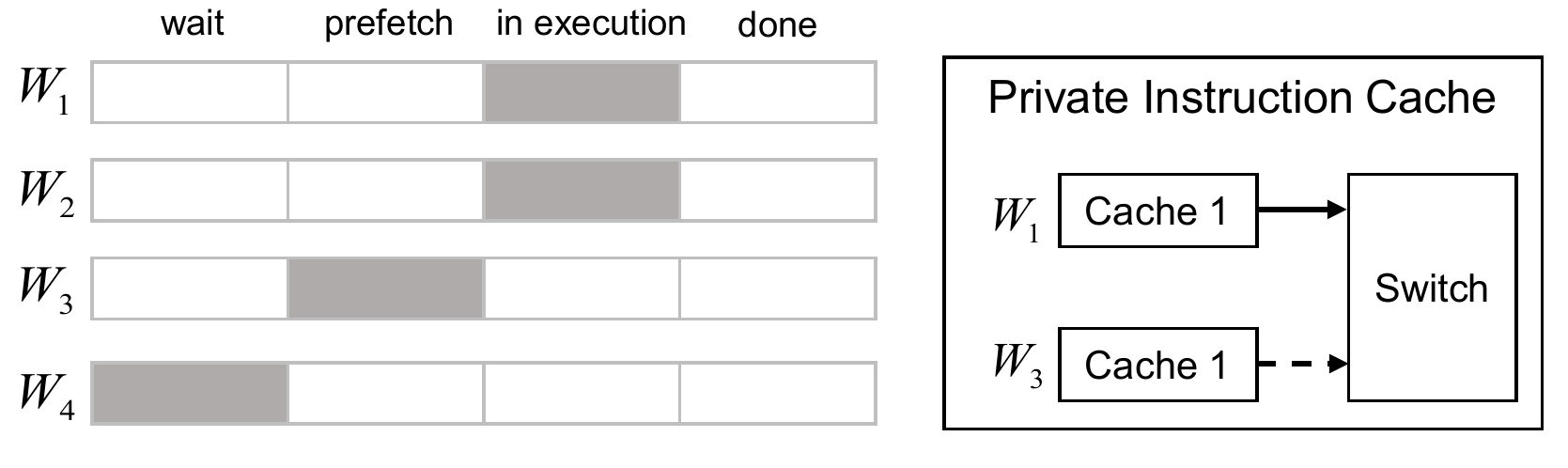}
         \caption{ }
         \label{fig:scheduler1}
     \end{subfigure}
     \hfill
     \begin{subfigure}[b]{0.48\textwidth}
         \centering
         \includegraphics[width=\textwidth]{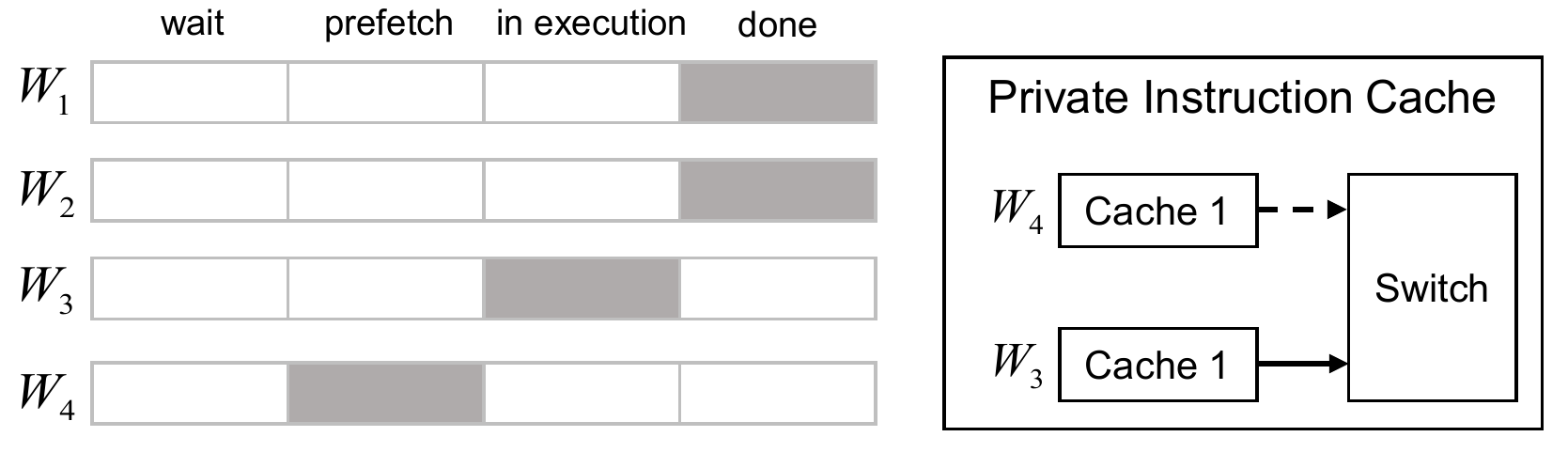}
         \caption{ \vspace{-3mm}}
         \label{fig:scheduler2}
     \end{subfigure}
    \setlength{\abovecaptionskip}{0pt}
    \caption{An example of the status register when the program block being executed is switched from $W_1$, $W_2$ to $W_3$: (a) the private instruction cache select the path of $W_1$, and $W_3$ is being prefetched into the second cache; (b) the private instruction switch to the second cache and $W_4$ start prefetching. \vspace{-3mm}}
    \label{fig:scheduler}
\end{figure}

    Therefore, we added an extra cache to the private instruction cache for each processor to prefetch the program block to be executed in the next step. At the same time, a \textit{prefetch} stage is added to the status register in scheduler. For instance, The initial state of the circuit module $W_3$ is "wait". When all dependent program blocks are in the "in execution" state, the scheduler will prefetch its corresponding instructions into the free cache and change the state of $W_3$ to "prefetch". When $W_1$ and $W_2$ are both finished, the scheduler will notify the processor to switch to the other cache. Its status is changed to "done" when the execution completes, which is used to indicate the scheduling of subsequent blocks. 
    The status flow in this example is shown in Figure \ref{fig:scheduler}.
    
    Fast block switching is achieved through this \textit{prefetch} scheduling method, which usually only consumes a few clock cycles to switch the cache path. This method helps to minimize the overhead caused sub-circuit switching at the cost of using additional caches.
    

\begin{figure*}[t]
\centering
\includegraphics[width=0.77\textwidth]{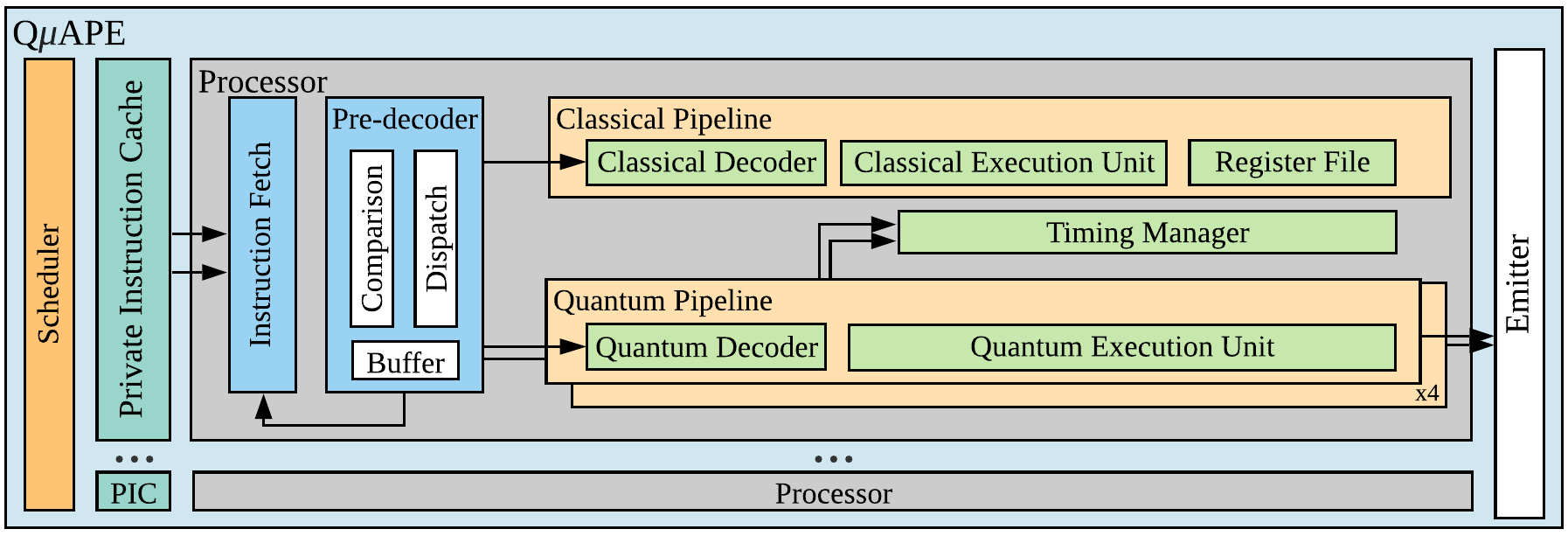}
\setlength{\abovecaptionskip}{3pt}
\caption{An overview of the quantum superscalar. \vspace{-4mm}}
\label{fig:superscalar}
\end{figure*}

\subsubsection{Processor}

    The previously described modules can be regarded as the control unit for determining the coordination of multiple processors. An overview of the processing unit is given below.

    (1) Register resource. Each processor has its own dedicated register for general purpose operations and comparison flags. Shared registers are also provided for all processors, which can be used for managing race condition and deadlock. 
    
    (2) Measurement result register. The measurement result register is written by the digital acquisition part. This register file can be shared by all processor because the processor can only read it.
    
    
    (3) Decoder and execution unit. The decoded information of classical and quantum instructions is sent to different execution units. A timing queue is used to buffer the generated timing information and is continuously read by the timing controller. The timing controller broadcast the label to all operation queues when the assigned timing is reached, thereby ensuring precise timing control.
    
    
    
    (4) Emitter. The last stage of the execution unit is to convert the operation for each qubit into a codeword sent to the low-level control electronics. For example, the microwave operation and flux operation for the same qubit need to be distributed to different analog channels due to the quantum processor setup. The previous modules of the control processor are independent of the analog control settings.

\subsection{Quantum Superscalar}
    
    In each processor of the above-mentioned multiprocessor architecture, we further employ quantum superscalar to exploit the parallelism of quantum instructions. Figure \ref{fig:superscalar} shows an example of a 4-way quantum superscalar structure. In this architecture, four instructions are fetched into the pre-decoder in one cycle and then dispatched to different pipelines for execution. 

    As mentioned in Section 5.1, the main goal of using the superscalar approach is to reduce the rapidly-growing QICES. Since introducing too much complexity in the comparison stage will limit the number of superscalar ways, our approach implements a \textit{parallel-until-classical} scheduling scheme.
    Potential timing hazards are prevented by performing comparison and recombination in the pre-decoder. A lookahead strategy is used to allow separate dispatch of different types of instructions, which helps to absorb the branching latency.

\subsubsection{Pre-decoder}
    
    The pre-decoder buffers the fetched instructions and distinguishes whether they are classical or quantum instructions, and then determines the dispatch of the instructions. In this mechanism, the following situations occur in the pre-decoder:
    
    \noindent \textbf{(1) All fetched instructions are quantum instructions:} Since the quantum operations of different timing labels should only enter the operation queue in serial, we only allow instructions with the same timing label to be issued simultaneously. Therefore, the timing dependency check of quantum instructions can be achieved by comparing their timing label with the first instruction. Quantum instructions with different timing label are buffered to be dispatched in the next cycle.
    
    The potential parallelism of quantum instructions may be disrupted if they are fetched in different cycles. For example, four parallel quantum instructions that are not fetched simultaneously cost two cycles for dispatching, which is not the ideal behavior. Therefore, we implement a series of buffers to store extra fetched instructions and synchronize instructions with the same timing label. In this case, parallel quantum instructions can be recombined for dispatching. The instruction fetch is stalled when all buffers are not empty. 
    
    \noindent \textbf{(2) Classical instructions in the fetched instructions:} Parallel quantum instructions are still scheduled as described above, while the classical instructions can only be dispatched in serial. The dispatch of classical instructions can be independent of quantum instructions. For instance, a branch instruction can be sent into the classical execution unit ahead of buffered quantum instructions. This lookahead mechanism helps to eliminate the latency caused by branch instructions and reduce CES. Therefore, we allow separate dispatch of quantum and classical instructions to exploit potential instruction-level parallelism.
    
    \vspace*{2mm}
    
    As shown in Figure \ref{fig:superscalar}, we only implement one classical execution unit for the parallel-until-classical scheme. In principle, this method can also be extended to allow parallel execution of classical instructions, but at the cost of limited scalability.
    For classical instructions, it is usually necessary to compare all pairs of instructions to check the dependency of input and output registers, which leads to $O(w^2)$ comparisons ($w$ is the issue width). 
    

\subsubsection{Classical and Quantum pipeline}

    To adapt to the separate dispatch of classical and quantum instructions, the decoder and execution unit as discussed in Section 5.2.4 is further divided. Each processor has one classical pipeline and multiple quantum pipelines, which decode and execute classical and quantum instructions respectively. 
    Only one timing controller is implemented in the processor to manage timing information, otherwise the timing control of different quantum instructions cannot be guaranteed. 
    

\subsection{Fast Context Switch}

    We first define simple feedback control as a special case that uses the measurement result of a single qubit to control a small number of quantum operations. Common applications of this type of control include active qubit reset and Bell state preparation \cite{riste2013deterministic}, which are useful primitive for various scenario. However, the simple feedback control is not suitable to be processed as a single program block because (1) frequently occurring parallel control flow can lead to a bloated number of processors, and (2) the small size of such program blocks complicates the scheduling mechanism by frequently requesting block switch. 
    
    To absorb the feedback control stall latency, we propose a mechanism for fast context switch. Since all the information of this kind of control can be put into one instruction for processing, the basic idea is to store the state of the system when the feedback control instructions are executed. Instead of stalling the pipeline, the processor can continue with instructions that are not related to this control.
 
    
    For instance, we can use a MRCE (Measurement Result Conditional Execution) instruction to indicate a simple feedback control process. An example of the syntax and encoding is given as follows:
    \vspace{-1mm}
    
    \begin{lstlisting}
        MRCE  qr0, q1, q_op0, q_op1
    \end{lstlisting}
    
    \vspace{-4mm}
    
    \begin{scriptsize}
    \begin{table}[h!]
      \centering
      \begin{tabular}{|l|l|l|l|l|}
        \hline
        \textbf{Opcode} & \textbf{q\_result\_addr} & \textbf{q\_target\_addr} & \textbf{op0} & \textbf{op1} \\
        \hline
      \end{tabular}
      \label{table:mrce}
    \end{table}
    \end{scriptsize}
    
    \vspace{-3mm}
    
    This example indicates that the operation to be performed on qubit 1 should be determined based on the measurement result of qubit 0. 
    The fast context switch is triggered when executing this instruction. Relevant information of this feedback control is stored, including the quantum operations and target qubits. The processor then continues to execute subsequent instructions until one of the following occurs: (1) The valid measurement result is returned, and the processor switches back to the MRCE instruction. (2) The pipeline reads an instruction about the stored qubits, and thus stalls due to the dependence of these quantum instructions. Since only a small amount of information needs to be stored, the processor can perform this switching with a short delay. This mechanism allows parallel execution of simple feedback control and quantum instructions that are irrelevant to this control. It also reduces the latency caused by conditional execution, thereby contributing to a lower CES.
    
    
    This method provides a flexible and fast execution of the simple feedback control. In principle, this mechanism can be extended to the circuit-level feedback control. However, the switching between large program blocks is much more complicated and time-consuming. In this work, we only allocate the parallel complex feedback control to different processors.
    
\section{Implementation}
    
    We implemented \sys{} using FPGA. Various implementations with different number of processors are prepared for evaluation. In order to validate the design, a final implementation is used to perform quantum experiments on the qubits. In addition, we also implemented (Arbitrary Waveform Generator) AWG and DAQ modules using customized hardware to achieve a complete control stack. This section introduces the implementation of our QCP and system.

\subsection{Quantum Control Processor}
    
    We use an Altera Stratix10 FPGA chip to implement the QCP, which contains the \sys{} and communication interface. Due to the relatively small size of current quantum programs, we directly use FPGA block RAM to implement all instruction memory and caches.
    The complete memory hierarchy is planned for future release.
    A series of measurement result registers are used to store the classical qubit states, which are obtained through demodulation, integration and  thresholding in the DAQ module. The \sys{} implements a block information table with 64 entries, each of which occupies 32 bits. Other presented modules in Section 5 are also implemented in the QCP. 
    The core fabric is clocked at 100MHz, and the communication interface uses a high-speed serial interface for low-latency data transmission.

\begin{figure}[h]
\includegraphics[width=0.35\textwidth]{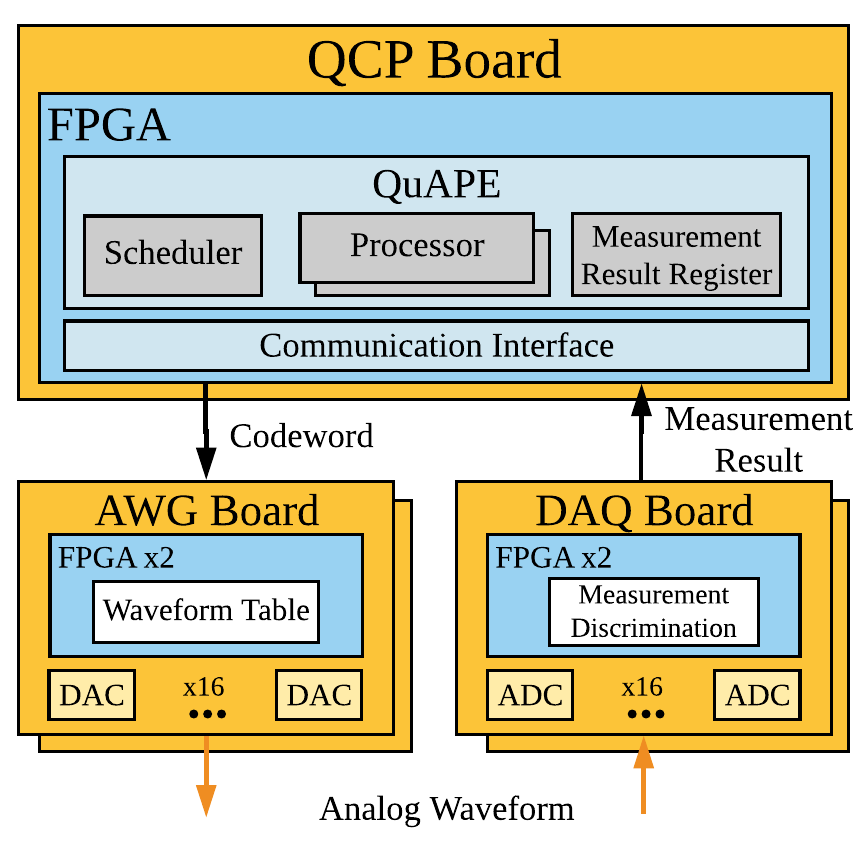}
\centering
\caption{System structure implementing the control stack. The orange arrows represent analog signals and the black arrows represent digital signals. \vspace{-3mm}}
\label{fig:system}
\end{figure}

\subsection{System}
    
    The schematic of the entire system is given in Figure \ref{fig:system}. The QCP sends codeword to AWGs to trigger the waveform generation, and receives measurement results from DAQs. The system consists of one QCP board, and multiple AWG and DAQ boards. In our implementation, each AWG has two aforementioned FPGAs, and each FPGA is connected to eight Digital-to-Analog Converters (DACs). The DAQ board has a similar structure, except that DACs are replaced with Analog-to-Digital Converters (ADCs). All AWGs and DAQs are also clocked at 100MHz. A backplane is implemented to provide connections from QCP board to all other boards. The backplane provides wiring up to 18 AWG boards and 2 DAQ boards, but not all of them are used in later experiments.
    
        
        
\section{Evaluation}
    
    In order to evaluate \sys{}, we perform some benchmark tests using only the QCP board. In this section, we investigate the impact of the number of processors on the program execution time, and evaluate the performance of the superscalar approach on multiple benchmarks. 
    
    \vspace{1.0mm}
    
\noindent \textbf{Circuit level parallelism evaluation:}

    \vspace{1.0mm}
    
\begin{figure}[h]
    \includegraphics[width=0.45\textwidth]{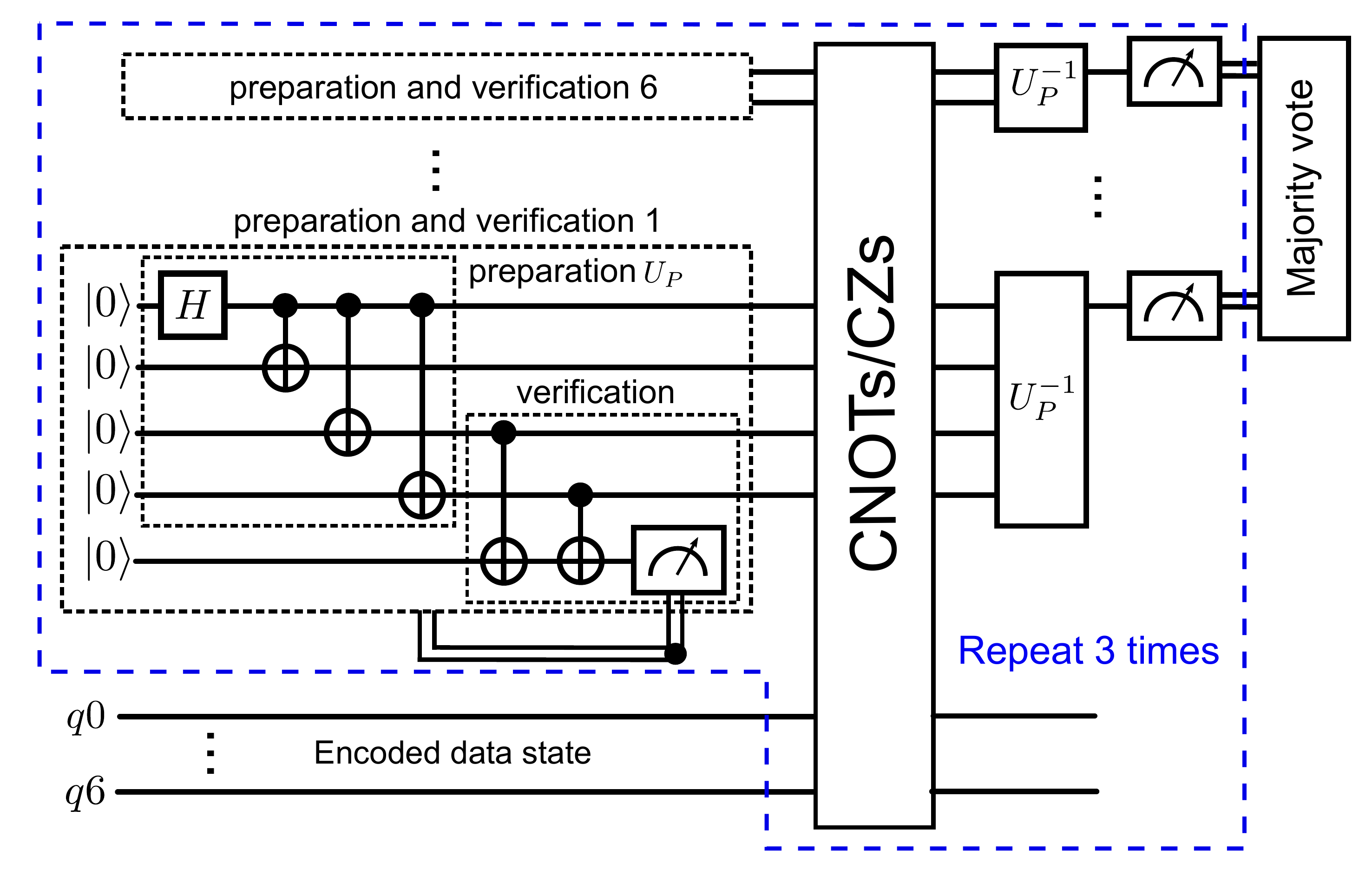}
    \centering
    \caption{Circuit diagram for fault-tolerant measurement. \vspace{-3mm}}
    \label{fig:ft_msmt}
    \end{figure}

    \noindent \textbf{Benchmark:} We first only focus on evaluate our multiprocessor architecture using a benchmark for standard Shor syndrome measurement circuits for 7-qubit Steane code \cite{steane1996error}. The circuit of this benchmark is given in Figure \ref{fig:ft_msmt}, where $\texttt{q}0-\texttt{q}6$ are encoded data qubit block. Each of its six stabilizer generators can be measured fault-tolerantly via bit-wise CNOT/CZ between these encoded data qubits block and 4-qubit ancilla cat state. Each cat state needs to be prepared followed by verifying certain parities because the preparation itself is not fault-tolerant. The whole circuit is repeated until the verification results are 0. This process needs to perform the verification for different parities for different ancilla blocks simultaneously. To establish reliable syndromes, one should take 3 times measurements followed by a majority vote.


    We use 37 qubits and assume all presented two-qubit connections are valid to avoid introducing unnecessary complexity in this microarchitecture-level test. 
    We also write a preliminary compiler to generate instructions for the evaluation and experiment. In this test, the decision of block division is made based on potential parallel sub-circuits. For example, the six stabilizers are assigned the same priority, but each one is treated as a separate block.
    The corresponding assembly program consists of 288 quantum instructions and 252 classical instructions, which indicates the existence of complex classical control in this benchmark. This quantum program is divided into 50 blocks with 15 different priorities to enable potential circuit-level parallelism. We use four different implementations with one, two, four and six processors to perform the benchmark test. In different implementations, the scheduler is only allowed to prefetch the first (first two/four/six) block(s) before the task starts. A pseudo random number generator is implemented in the FPGA to generate measurement results for testing. The feedback control latency is measured to be approximately 450ns. We take the uniprocessor implementation as the baseline design and compared it with other three designs. 
    
    \noindent \textbf{Evaluation metric:} The latency of feedback control typically requires hundreds of nanoseconds, and it is significant compared to the gate time (usually 20~40ns). Therefore, we use execution time and the achieved speedup as the metric for CLP exploitation, because the classical control part limits the execution time. In the near term, the prolonged execution time will significantly affect the fidelity due to decoherence errors. In the long term, with the continuously improving quality of the quantum chip, speedup will become a dominant factor for realizing an efficient quantum computer.
    

    \begin{figure}[h]
    \centering
    \begin{subfigure}{0.49\columnwidth}
         \centering
         \includegraphics[width=\columnwidth]{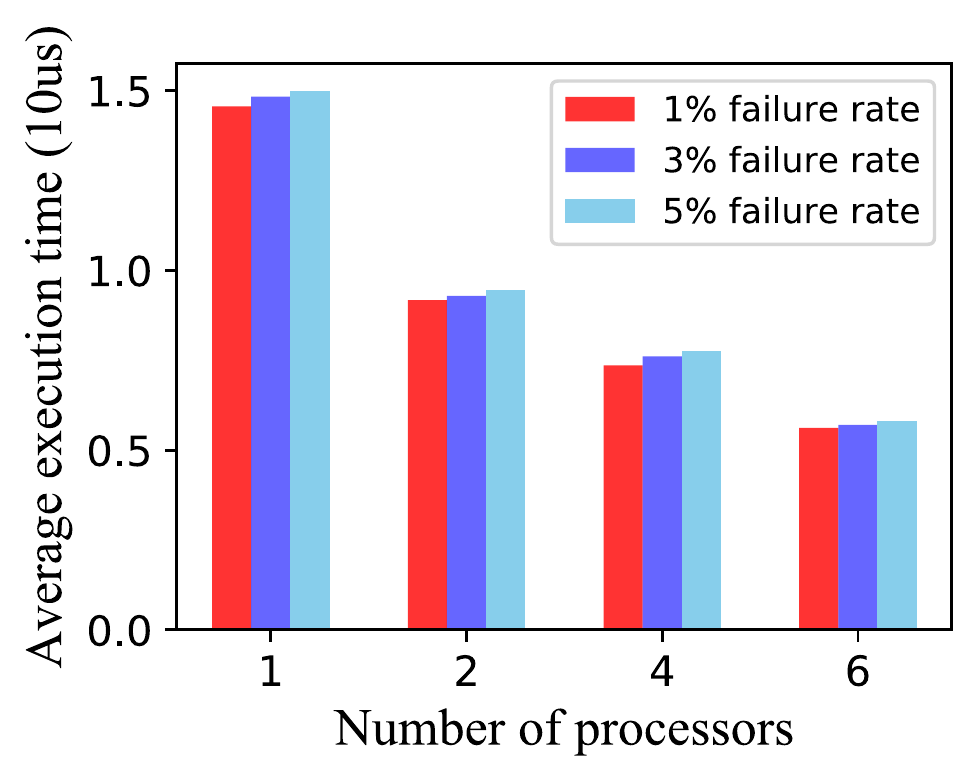}
         \caption{ }
         \label{fig:clp1}
     \end{subfigure}%
     \begin{subfigure}{0.49\columnwidth}
         \centering
         \includegraphics[width=\columnwidth]{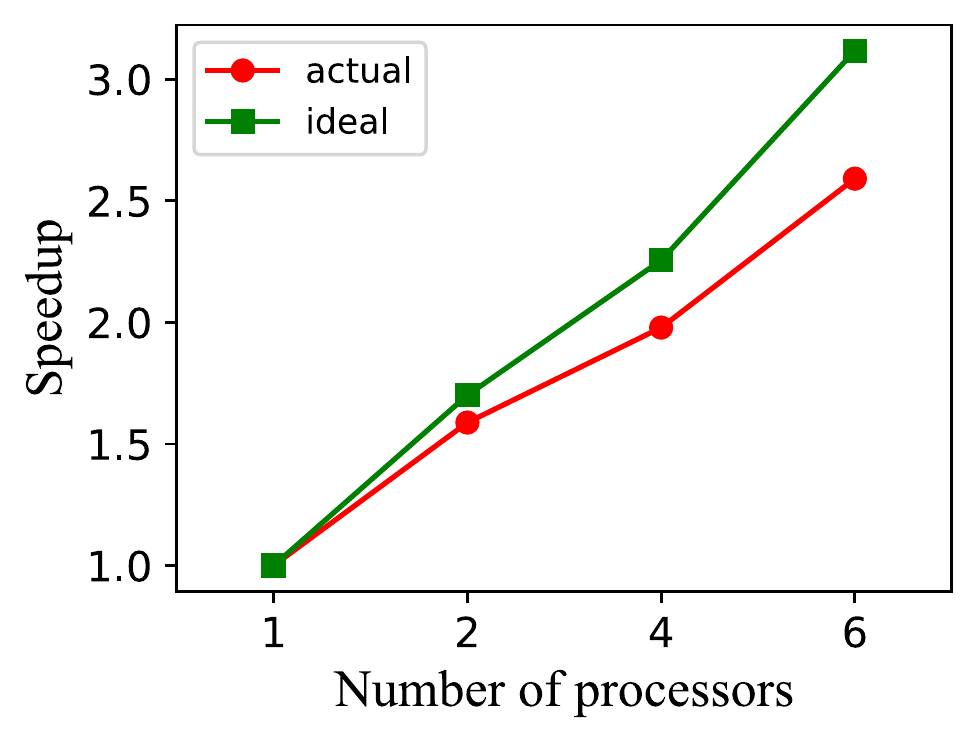}
         \caption{ }
         \label{fig:clp2}
     \end{subfigure}%
        \setlength{\abovecaptionskip}{3pt}
        \caption{Results of benchmark tests performed using different number of processors: (a) average execution time with three different failure rates; (b) the actual and ideal speedup. \vspace{-3mm}}
        \label{fig:clp result}
    \end{figure}

    \noindent \textbf{Results:} The results are shown in Figure \ref{fig:clp result}. Figure \ref{fig:clp1} shows the execution time of different implementations, and each result is obtained by averaging the results of 1000 executions. The failure rate indicates the probability of failure of the preparation step, which leads to a longer execution time. The actual speedup is obtained by comparing the observed execution time with the baseline design. As a comparison, we also calculate the theoretical speedup in Figure \ref{fig:clp2} by assuming that all block scheduling and allocation can be completed without taking any clock cycles. 
    
    
    The difference between the actual speedup and theoretical speedup mainly due to: 1) the scheduling response time spent in the scheduler, and 2) the allocation time for fetching instructions into the private cache. Our prefetch mechanism works well by itself, though some program blocks remains unaccelerated due to the frequent block switching. In this test, the smallest block only has four instructions. This demonstrate that dividing program into fine-grained blocks can even have negative impact, because the scheduler cannot respond to overwhelming concurrent requests. Overall, our six-processor implementation achieved a \textbf{2.59$\times$} speedup compared to the uniprocessor. This substantial improvement can significantly reduce the accumulate quantum errors during computation.
    
\begin{figure}[h]
    \includegraphics[width=0.45\textwidth]{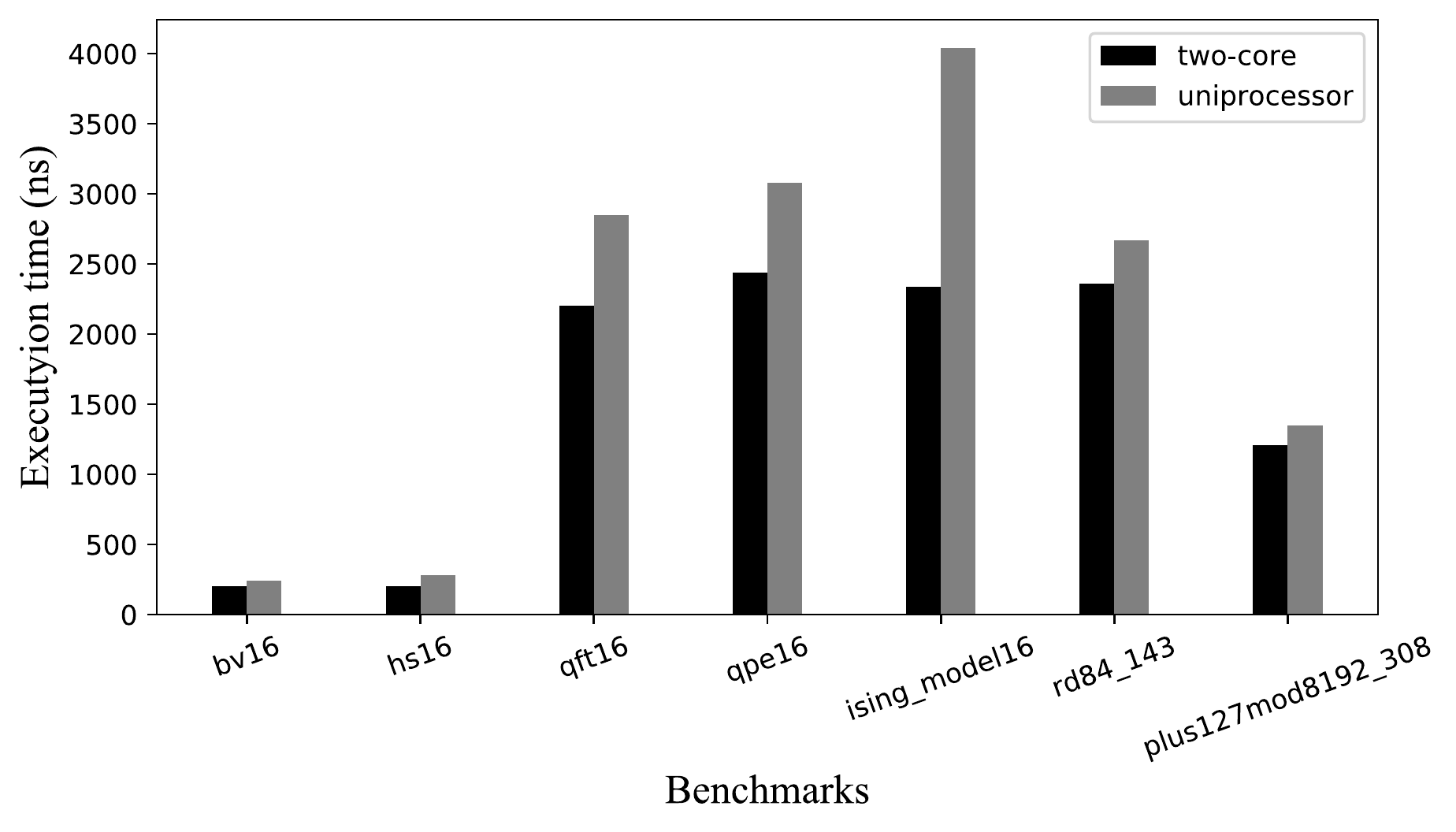}
    \centering
    \caption{Execution time for various benchmarks on two-core and uniprocessor.} \vspace{-3mm}
    \label{fig:quantum}
    \end{figure}

    
    To conduct a more comprehensive evaluation of the multiprocessor architecture, we also selected 7 different benchmarks from Qiskit \cite{Qiskit}, ScaffCC \cite{scaffcc}, and RevLib \cite{WGT+:2008} for testing. We measure the execution time of each benchmark to calculate the speedup that a two-core implementation can achieve compared to the uniprocessor implementation. The results are shown in Figure \ref{fig:quantum}.
    
    In these tests, we simply divide the part of the program with parallel operations into two blocks, each corresponding to half of the qubits. Overall, the two-core implementation can achieve an average \textbf{1.30$\times$} speedup of these benchmarks. This result proves that the multiprocessor architecture can help a variety of applications to obtain performance improvements, thereby alleviating the limitations of the classic control part on program execution time.

    \vspace{1.0mm}
    
\noindent \textbf{Quantum operation level parallelism evaluation:}

    \vspace{1.0mm}

\begin{figure}[h]
    \includegraphics[width=0.45\textwidth]{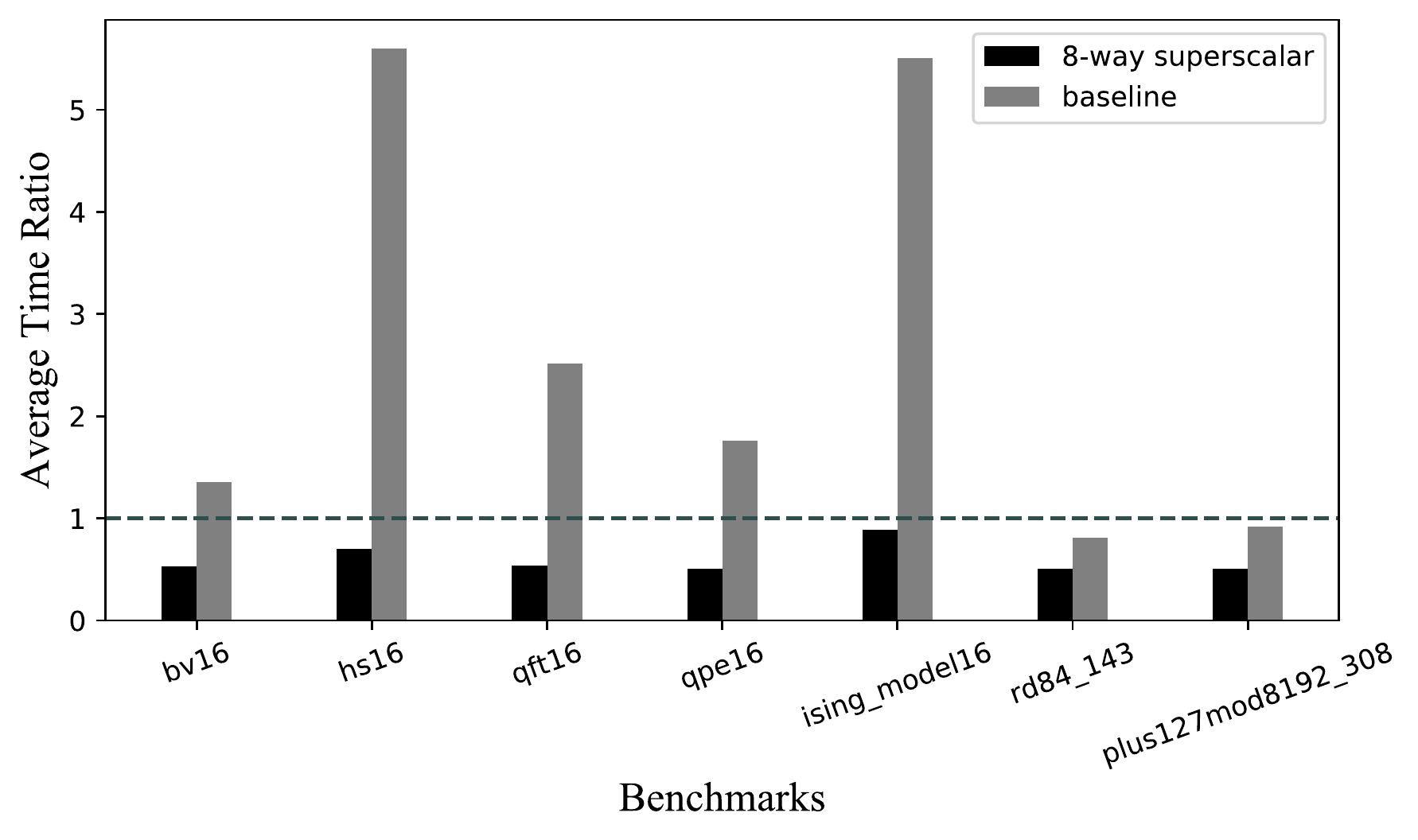}
    \centering
    \caption{Results of average TR for the 8-way superscalar baseline design. The dotted line represents $TR = 1$. \vspace{-3mm}}
    \label{fig:qolp}
    \end{figure}

    We implemented an 8-way superscalar architecture on the QCP board. We use the same benchmarks to demonstrate the performance of our quantum superscalar architecture, while a scalar processor is used as the baseline design.
    We measured the number of clock cycles spent for each circuit step in the QCP. The TR is then calculated by setting the clock time to 10ns and the gate time to 20ns. We calculated the average TR for a direct comparison. Figure \ref{fig:qolp} shows the results of the 8-way superscalar and baseline design on each benchmark.
    
    Overall, our superscalar approach achieve an average \textbf{4.04$\times$} reduction in the average TR of these seven benchmarks. An \textbf{8.00$\times$} reduction is observed in the result of the \textit{hs16} benchmark, which achieves theoretical upper bound. Due to the relatively small degree of parallelism that can be exploited, only \textbf{1.60$\times$} improvement is achieved for \textit{rd84\_143}. For the last two benchmarks, the average TR of the baseline design is less than 1. However, their maximum TR are 4.5 and 9, respectively. As a comparison, our design reached $TR \leq 1$ for all seven benchmarks. 
    
    
    We also verified the fast context switch in \sys{}. We used a program with an active qubit reset and a randomized benchmarking (RB) for testing. The RB instructions can be executed correctly when the active qubit reset waits for its measurement result. In this implementation, we measured that it takes three clock cycles to switch the context of the simple feedback control.

\section{Experiments}
The \sys{} processor implemented for the experiment targets a 10-qubit one-dimension superconducting chip, which requires 38 analog channels for control and readout in our experimental setup. The connection information between the analog devices and the quantum chip is hard-coded in \sys{} to provide the appropriate control signals. Details of our experimental set-up can be found in \cite{zhou2021rapid}.
    
\begin{figure}
\centering
\includegraphics[width=0.45\textwidth]{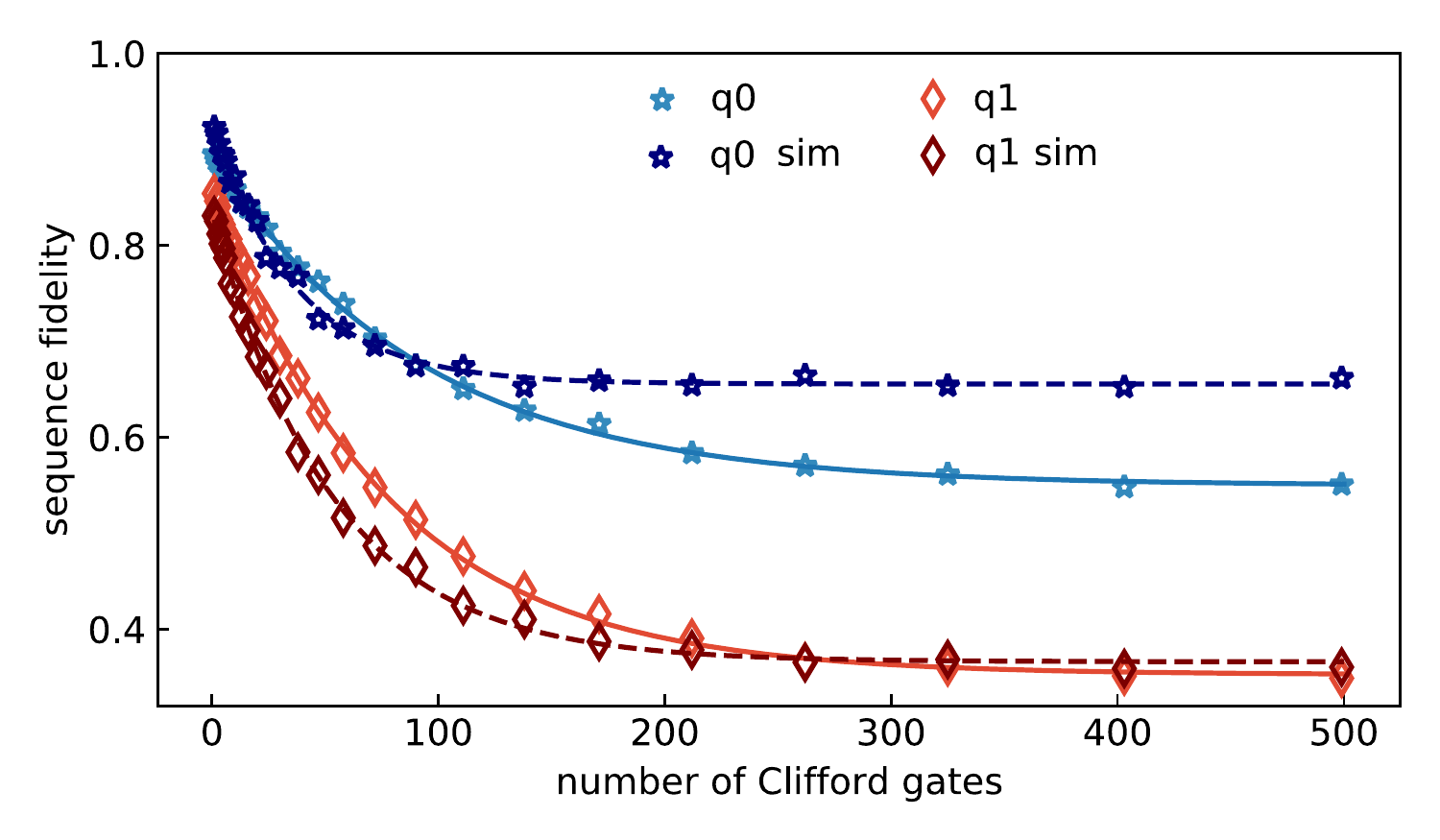}
\caption{Experiments for RB and simRB on two qubits (q0 and q1). Symbols are experimental data, and solid (dashed) curves represent the fitting result for RB (simRB). \vspace{-3mm}}
\label{fig:fig_rb_expt}
\end{figure}

We perform the RB experiments both individually \cite{knill2008randomized} and simultaneously\cite{gambetta2012characterization} on a pair of qubits (q0 and q1) with our 10-qubit QPU. As illustrated in Figure \ref{fig:fig_rb_expt}, the blue star symbols represent the experimental data for q0, and the red thin diamond ones for q1 respectively. The individual RB experiments showed in light blue (red) color for q0 (q1) are performed as a reference, and we extract the average fidelity of single qubit gates are 99.5\% (99.4\%) for q0 (q1). The data for simultaneous randomized benchmarking (simRB) is shown in dark red (blue) for q0 (q1). The mean gate fidelity is 98.7\% and 99.1\% for q0 and q1, respectively. The lower fidelities comparing with the individual RB is accounted for the inevitable ZZ interaction between the qubits.

This result demonstrates that the multiprocessor and quantum superscalar architectures in \sys{} can work properly. This experiment shows that \sys{} is capable of simultaneously issuing quantum operations to different qubits, which can be used to control more qubits. However, our current experimental setup is restricted by the read-out device. In this experiment, the simRB is applied to two qubits.

\section{Related Work}

    \noindent \textbf{Comparison with QuMA\_v2\cite{fu2019eqasm}:} Table 2 shows the characteristics of QuMA\_v2 and \sys{}. QuMA\_v2 also has a centralized memory architecture and supports various control flow by executing eQASM instructions. QuMA\_v2 currently does not support the direct exploitation of CLP, making it difficult to handle applications such as the fault tolerant syndrome measurement. In the tests of multiprocessor (see Section 7), the uniprocessor implementation can be regarded as QuMA\_v2.
    
\begin{scriptsize}
\begin{table}[h!]
  \centering
  \begin{tabular}{|l|l|l|l|}
    \hline
    & \textbf{\sys{}} & \textbf{QuMA\_v2,}\\
    &  & \textbf{HPCA 2019 \cite{fu2019eqasm}}\\
    \hline
    \hline
    Target technology & Superconducting & Superconducting\\
    \hline
    Memory& Centralized & Centralized\\
    architecture &  &\\
    \hline
    CLP & Multiprocessor & N/A\\
    \hline
    QOLP & Superscalar & VLIW, SOMQ \\
    \hline
    Feedback control & Supported & Supported\\
    \hline
  \end{tabular}
  \setlength{\abovecaptionskip}{5pt}
  \caption{Comparison with QuMA\_v2 \vspace{-3mm}}
  \label{table:evaluation}
\end{table}
\end{scriptsize}
  
    Regarding QOLP, we choose superscalar over the very-long-instruction-word (VLIW) approach in QuMA\_v2 mainly due to the following three reasons: First, the length of a single instruction can remain unchanged when implementing more execution units, thereby ensuring a fixed-length QISA design. Such RISC-fashion instruction set enables a compact and efficient microarchitecture implementation. Second, the amount of inserted QNOPs in the VLIW bundle will lead to additional program size. Third, the separate dispatch of classical and quantum instructions in the superscalar approach absorbs the potential delay caused by branch instructions.  
    
    QuMA\_v2 also adopts a single-operation-multiple-qubit (SOMQ) method to apply a single operation on multiple qubits simultaneously. This technology can reduce the QICES in Equation \ref{eq:CES} and is beneficial for achieving a lower CES. However, the analysis in \cite{fu2019eqasm} assumes that the QCP can always provide all the target qubit (pair) list in time, which is difficult to achieve in some quantum experiments, such as the quantum random circuit sampling \cite{arute2019quantum}.
    
    
    
    
    
    \vspace*{1.5mm}
    
    \noindent \textbf{Other related work:} The APS2 system proposed in \cite{ryan2017hardware} has a distributed architecture, which means that a single quantum program needs to be translated into multiple assembly programs to run on different modules. Tannu \textit{et al.} proposed QuEST \cite{tannu2017taming}, a quantum control processor architecture that can tame the instruction bandwidth problem mainly caused by quantum error correction code instructions. 
    A NISQ quantum computer simulator including the classical control system can help the microarchitecture research to perform a more comprehensive simulation. Several QC simulators have been proposed, such as QPDO \cite{riesebos2017pauli} and SANQ \cite{li2019sanq}, which can be extended to achieve architectural simulation like traditional simulators, e.g., GEM5 \cite{binkert2011gem5}. 
    
    Recent works on compilers \cite{shi2019optimized, abhari2012scaffold, li2019tackling} has also paved the way for full-stack quantum computers. In this research, we only wrote a preliminary compiler to generate instructions for evaluation and experiment. As more advanced compilation techniques can help find more opportunities for parallelism exploitation, we can conduct more in-depth explorations based on our microarchitecture-level proposal in the future, e.g. block division methods and trade-offs between parallelism and cross-talk.
\section{Conclusion}

    In this work, we clarified the requirements for exploiting CLP and QOLP in the control microarchitecture. To tackle the limited operation issue rate of quantum control microarchitecture, we propose \sys{} to exploit different levels of parallelism. In this work, we propose three mechanisms: multiprocessor, quantum superscalar, and fast context switch for simple feedback control. A multiprocessor architecture can achieve concurrent processing of multiple program blocks. Quantum superscalar architecture reduces the CES in an effective way. The context switch mechanism contributes to a fast and flexible feedback control, which helps the quantum computing to benefit from dynamic quantum circuits. 
    
    We implemented a \sys{} prototype on FPGA. Multiple benchmarks are tested to demonstrate the capability of our microarchitecture. The design is validated by performing a simRB experiment on a superconducting QPU. In conclusion, our proposal offers insights for building a scalable quantum control microarchitecture for larger NISQ systems.

\section*{ACKNOWLEDGMENTS}
We thank the anonymous reviewers for their insightful feedback. We also thank Sainan Huai, Yu Zhou for contributions to the experimental set-up, Dongning Zheng for fabricating the quantum chip. This work is funded in part by Key-Area Research and Development Program of Guangdong Province, under grant 2020B0303030002.

\bibliographystyle{IEEEtranS}
\bibliography{refs}

\end{document}